\documentclass[conference]{IEEEtran}
\IEEEoverridecommandlockouts

\newcommand{\TT}{\mathsf{T}}
\newcommand{\HH}{\mathsf{H}}

\newcommand{\av}{{\bf a}}

\newcommand{\cv}{{\bf c}}

\newcommand{\fv}{{\bf f}}

\newcommand{\hv}{{\bf h}}

\newcommand{\nv}{{\bf n}}

\newcommand{\pv}{{\bf p}}

\newcommand{\sv}{{\bf s}}

\newcommand{\wv}{{\bf w}}
\newcommand{\vv}{{\bf v}}

\newcommand{\yv}{{\bf y}}
\newcommand{\zv}{{\bf z}}

\newcommand{\Bm}{{\bf B}}

\newcommand{\Dm}{{\bf D}}

\newcommand{\Fm}{{\bf F}}
\newcommand{\Gm}{{\bf G}}
\newcommand{\Hm}{{\bf H}}

\newcommand{\Mm}{{\bf M}}
\newcommand{\Nm}{{\bf N}}

\newcommand{\Sm}{{\bf S}}

\newcommand{\Wm}{{\bf W}}

\newcommand{\Ym}{{\bf Y}}

\newcommand{\Bt}{{\mathsf B}}

\newcommand{\Ft}{{\mathsf F}}

\newcommand{\Ut}{{\mathsf U}}

\newcommand{\phiv}{\hbox{\boldmath$\phi$}}

\newcommand{\thetav}{\hbox{$\boldsymbol\theta$}}

\usepackage{acronym}

\acrodef{5G}{the fifth generation}
\acrodef{MIMO}{multiple-input multiple-output}
\acrodef{MISO}{multiple-input single-output}
\acrodef{RF}{radio frequency}
\acrodef{BS}{base station}
\acrodef{UE}{user equipment}
\acrodef{LoS}{line-of-sight}
\acrodef{NLoS}{non-line-of-sight}
\acrodef{AoA}{angle-of-arrival}
\acrodef{AoD}{angle-of-departure}
\acrodef{UPA}{uniform planar array}
\acrodef{ARV}{array response vector}
\acrodef{CGV}{channel gain vector}
\acrodef{AGV}{antenna gain vector}
\acrodef{EM}{electromagnetic}
\acrodef{MA}{movable antenna}
\acrodef{3D}{three-dimensional}
\acrodef{LS}{least-squares}
\acrodef{AWGN}{additive white Gaussian noise}
\acrodef{ERA}{electromagnetically reconfigurable antenna}
\acrodef{LMMSE}{linear minimum mean-square error}
\acrodef{SNR}{signal-to-noise ratio}
\acrodef{DNN}{deep neural network}
\acrodef{MLP}{multi-layer perceptron}
\acrodef{LSTM}{long short-term memory}

\usepackage{pgfplots}       % High-quality plots using TikZ
\pgfplotsset{compat=1.18}   % Specify PGFPlots compatibility mode
\usepackage{xcolor}         % Define and manage custom colors

\definecolor{profBlue}{RGB}{0,91,150}        % Deep muted blue
\definecolor{profCyan}{RGB}{0,150,170}       % Calm teal cyan
\definecolor{profPurple}{RGB}{120,60,180}    % Muted violet
\definecolor{profOrange}{RGB}{200,90,10}     % Deep amber orange
\definecolor{profGray}{RGB}{90,90,90}        % Neutral dark gray
\definecolor{profGreen}{RGB}{0,130,90}       % Forest teal green
\definecolor{profRed}{RGB}{180,60,50}        % Dark scientific red

\usepackage{graphicx}       % Include images
\usepackage{subcaption}     % Subfigures (side-by-side figures)
\usepackage[bottom=0.9in,top=0.9in,left=0.9in,right=0.9in]{geometry}  % Adjust page margins
\usepackage[font=footnotesize]{caption} % Customize caption font size

\usepackage{cite}           % Clean citation formatting
\usepackage{amsmath,amssymb,amsfonts} % Math symbols and fonts
\usepackage{algorithmic}    % Pseudocode for algorithms
\usepackage{textcomp}       % Additional text symbols

\def\BibTeX{{\rm B\kern-.05em{\sc i\kern-.025em b}\kern-.08em
    T\kern-.1667em\lower.7ex\hbox{E}\kern-.125emX}}

\begin{document}
\title{Model-Free Channel Estimation for Massive MIMO: A Channel Charting-Inspired Approach
\author{
Pinjun Zheng, Md. Jahangir Hossain, and Anas Chaaban\\
\textit{School of Engineering, University of British Columbia, Kelowna, Canada}\\
(Email: pinjun.zheng@ubc.ca) 
} 
}

\maketitle

\begin{abstract}
Channel estimation is fundamental to wireless communications, yet it becomes increasingly challenging in massive multiple-input multiple-output (MIMO) systems where base stations employ hundreds of antennas. Traditional least-squares methods require prohibitive pilot overhead that scales with antenna count, while sparse estimation methods depend on precise channel models that may not always be practical. This paper proposes a model-free approach combining deep autoencoders and LSTM networks. The method first learns low-dimensional channel representations preserving temporal correlation through augmenting a channel charting-inspired loss function, then tracks these features to recover full channel information from limited pilots. Simulation results using ray-tracing datasets show that the proposed approach achieves up to 9 dB improvement in normalized mean square error compared to the least-squares methods under ill-conditioned scenarios, while maintaining scalability across MIMO configurations.
\end{abstract}


\section{Introduction}
Accurate channel estimation is essential for reliable wireless data transmission. In massive \ac{MIMO} systems~\cite{Larsson2014Massive}, traditional methods face challenges as large antenna arrays result in high-dimensional channel matrices, thus requiring substantial pilot overhead for accurate estimation. When pilot signals are insufficient, the estimation problem becomes ill-conditioned. Sparse channel estimation methods can help in this case by reducing the number of unknowns by exploiting inherent channel structure~\cite{Heath2016Overview}. However, they rely heavily on accurate a priori channel models that may not always be available or precise in practice.

Data-driven deep learning approaches offer an alternative that does not require explicit channel models~\cite{Belgiovine2021Deep}. Early works include \ac{DNN}-based channel estimation, which directly learns the mapping from received pilot signals to channels~\cite{Hu2021Deep}. More recent advances exploit the spatial and temporal correlations in massive \ac{MIMO} channels and leverage transformer architectures to capture long-range dependencies~\cite{Zhou2024Pay}. However, these methods typically require training large-scale neural networks whose complexity scales with the number of antennas, making them computationally intensive for massive \ac{MIMO} deployments. Hybrid approaches that combine model-based and learning-based techniques have also been investigated~\cite{Ma2021Model}, but they still depend on some prior knowledge of the channel structure. Moreover, channel charting has emerged as a technique for learning low-dimensional embeddings from channels that preserve spatial relationships~\cite{Studer2018Channel}, but it mainly targets localization rather than channel estimation. Despite these advances, there remains a gap in developing scalable model-free approaches for massive \ac{MIMO} channel estimation that maintain low complexity during both training and inference.

To fill this gap, this paper proposes a method that first trains an autoencoder to map high-dimensional channel matrices to a low-dimensional latent space while preserving temporal correlation. This is realized through a carefully designed distance similarity loss function, which is inspired by channel charting~\cite{Studer2018Channel}. Subsequently, an \ac{LSTM} network tracks the temporal evolution of these latent states and recovers complete channels from limited pilot observations. A key advantage over existing data-driven approaches is the decomposition of the problem into static autoencoder and dynamic \ac{LSTM} training, where channel dimensionality only affects the easily trainable autoencoder while the complex \ac{LSTM} tracking operates on low-dimensional latent states independent of the channel scale. The success of our approach hinges on the autoencoder training design that ensures the learned latent space preserves temporal correlation, which is a crucial property for effective state tracking.

\section{Problem Description and Motivation}
We consider a channel estimation problem where a \ac{UE} equipped with $N_\mathsf{U}$ antennas transmits known pilot signals to a \ac{BS} with $N_\mathsf{B}$ antennas to estimate the uplink wireless channel. 

\subsection{Signal Model}
Let $\Hm\in\mathbb{C}^{N_\mathsf{B}\times N_\mathsf{U}}$ denote the uplink channel to be estimated. To simplify the case, we assume both the \ac{UE} and \ac{BS} are equipped with a single \ac{RF} chain and an analog beamforming architecture implemented via a \ac{RF} phase shift network. When estimating the channel, the \ac{UE} transmit to the \ac{BS} $M_\mathsf{U}$ known pilot symbols $s_i,\,,i=1,2,\dots,M_\mathsf{U}$, each through a precoder $\fv_i\in\mathbb{C}^{N_\mathsf{U}}$. For each symbol transmitted from the \ac{UE}, the \ac{BS} records it $M_\mathsf{B}$ times through different combiners $\wv_j\in\mathbb{C}^{N_\mathsf{B}},\, j=1,2,\dots,M_\mathsf{B}$. Define $\Fm=[\fv_1,\fv_2,\dots,\fv_{M_\mathsf{U}}]\in\mathbb{C}^{N_\mathsf{U}\times M_\Ut}$ and $\Wm=[\wv_1,\wv_2,\dots,\wv_{M_\mathsf{B}}]\in\mathbb{C}^{N_\mathsf{B}\times M_\Bt}$. We collect all the received signals at the \ac{BS}, and denote the matrix of the received signals as
\begin{equation}\label{eq:SigModel}
	\Ym = \Wm^\HH \Hm \Fm \Sm + \Nm \in\mathbb{C}^{M_\Bt\times M_\Ut},
\end{equation}
where $\Sm=\mathrm{diag}\{s_1,s_2,\dots,s_{M_\mathsf{U}}\}$ and $\Nm$ denotes an additive noise. To simplify notations, we define $\Gm=\Fm\Sm$ and thus~\eqref{eq:SigModel} becomes $\Ym = \Wm^\HH \Hm \Gm + \Nm$. Note that the total number of observations, i.e., $M_\Bt M_\Ut$, reflects the \emph{signaling overhead}. 

The channel estimation problem refers to estimating $\Hm$ based on $\Ym$. The following subsections recap two types of predominant solutions existing in the literature. 

\subsection{LS-Based Channel Estimation}\label{sec:LS}
Without any knowledge of channel structure or statistics, one can estimate the unknown channel by solving the following \ac{LS} problem: 
\begin{align}\label{eq:LS}
	\hat{\Hm}_\mathsf{LS} &= \arg\min_{\Hm}\ \|\Ym - \Wm^\HH \Hm \Gm\|_\Ft^2\\
	&= \arg\min_{\Hm}\ \|\mathrm{vec}(\Ym) - (\Gm^\TT\otimes\Wm^\HH) \mathrm{vec}(\Hm) \|_2^2, \notag
\end{align}
where $\mathrm{vec}(\cdot)$ denotes the column-wise vectorization of a matrix, and $\otimes$ stands for the Kronecker product. 

It is trivial to see that if $\Mm\triangleq \Gm^\TT\otimes\Wm^\HH\in\mathbb{C}^{M_\Bt M_\Ut \!\times\! N_\Bt N_\Ut}$ is full column rank, \eqref{eq:LS} has a unique closed-form solution given by 
\begin{equation}\label{eq:LSsolu}
	\mathrm{vec}(\hat{\Hm}_\mathsf{LS}) = (\Mm^\HH\Mm)^{-1}\Mm^\HH\mathrm{vec}(\Ym).
\end{equation}
However, holding this uniqueness condition is challenging in massive \ac{MIMO} systems, as usually $N_\Bt$ is large (e.g., $N_\mathsf{B} = 100$ in~\cite{Larsson2014Massive}) while the signaling overhead $M_\Bt M_\Ut$ is limited. When $M_\Bt M_\Ut < N_\Bt N_\Ut$, $\Mm^\HH\Mm$ in~\eqref{eq:LSsolu} is non-invertible and \eqref{eq:LS} has infinitely many solutions. In general, we can choose the one with the minimum norm, which is given by $\mathrm{vec}(\hat{\Hm}_\mathsf{LS}) = \Mm^\dag \mathrm{vec}(\Ym)$,
where $(\cdot)^\dag$ denotes the Moore–Penrose pseudoinverse. However, this minimum-norm solution may deviate significantly from the true channel.

\subsection{Sparse Channel Estimation}
Alternatively, channel estimation can be performed by exploiting the inherent structure of the channel. In high-frequency communications, such as mmWave and THz bands, wireless channels exhibit spatial sparsity in the angular domain (or far-field beamspace)~\cite{Heath2016Overview,Tarboush2021TeraMIMO}. For example, we can express the channel in the frequency domain as a superposition of multipath components as
\begin{align}\label{eq:Hexpression}
	\Hm = \sqrt{\frac{N_\Bt N_\Ut}{L}}\sum_{\ell=1}^L \rho_\ell \av_\Bt(\phiv_\ell) \av_\Ut^\HH(\thetav_\ell),
\end{align} 
where $L$ is the total number of propagation paths, $\rho_\ell$ denotes the complex channel gain, and $\av_\Bt(\phiv_\ell)$ and $\av_\Ut(\thetav_\ell)$ are array response vectors corresponding to the $\ell^{\text{th}}$ path at the \ac{BS} and \ac{UE}, respectively. Here, $\phiv_\ell$ denotes the \ac{AoA} at the \ac{BS} and $\thetav_\ell$ denotes the \ac{AoD} at the \ac{UE}. The detailed expression of these array response vectors in the 3D space can be found in, e.g.,~\cite[Eq.~(2)]{Zheng2025Mutual}.

This sparsity nature can be leveraged to facilitate channel estimation, as the channel matrix $\Hm$ is fully characterized by only a few parameters. Estimating these low-dimensional parameters $\{\rho_\ell, \phiv_\ell, \thetav_\ell\}_{\ell=1}^L$ is sufficient to reconstruct the entire channel. The estimation of these channel parameters based on the received signals can be realized using techniques such as compressed sensing~\cite{Alkhateeb2014Channel} and tensor decomposition~\cite{Zheng2024JrCUP}. 

\subsection{Motivation of This Work}
While both \ac{LS}-based and sparse channel estimation methods are well established, they each suffer from inherent limitations. \ac{LS}-based estimators require prohibitive signaling overhead for massive \ac{MIMO} systems due to the large channel. Sparse methods, while more efficient, rely on accurate channel models that may not always be available. For example, channels in low-frequency bands exhibit much weaker sparsity. Even at mmWave frequencies, non-ideal factors like spatial non-stationarity~\cite{Yuan2023Spatial} can lead to severe model mismatch, ultimately degrading the sparse estimation results.

In light of the aforementioned limitations, this paper aims to develop a channel estimation method that (i)~operates without assuming any explicit channel structure while (ii)~achieving effective performance under low signaling overhead constraints. The proposed method is primarily based on \emph{deep learning} techniques.

\section{Methodology Overview}
Before developing our method, we present a few considerations to illustrate the core ideas behind the proposed approach. 
\begin{itemize}
    \item \textbf{In static scenarios, structure plays a crucial role in overcoming ill-conditioning.} Sparse channel estimation methods work effectively because they exploit the inherent structure of the channel. The structural model significantly reduces the number of unknowns to be estimated, thereby alleviating ill-conditioning caused by insufficient observations. This remains true even when the model is inaccurate or unknown, i.e., the channel matrix is determined by a few \textit{low-dimensional features}, which can be learned by a deep neural network implicitly.
    \item \textbf{In dynamic scenarios, temporal correlation provides an additional means to alleviate ill-conditioning.} Typically, channel estimation is performed once per channel coherence interval. While the channel matrix itself may vary significantly across coherence intervals, some inherent features of the channel (e.g., the \acp{AoD} and \acp{AoA} in model~\eqref{eq:Hexpression}) vary slowly and smoothly over time in most realistic scenarios where the user is not moving rapidly. By exploiting this \emph{temporal correlation}, observations across multiple coherence intervals can be used to jointly track the inherent features, rather than estimating each channel matrix independently. 
\end{itemize}

Based on the above considerations, the proposed method consists of two main steps: (i)~learning a low-dimensional representation of the channel that preserves the temporal correlation property, and (ii)~tracking this latent representation using observations from multiple intervals to recover the full channel matrix. The following sections detail these two steps.

\section{Step 1: Latent Channel Representation}\label{sec:autoencoder}
\begin{figure}[t]
  \centering
  \includegraphics[width=\linewidth]{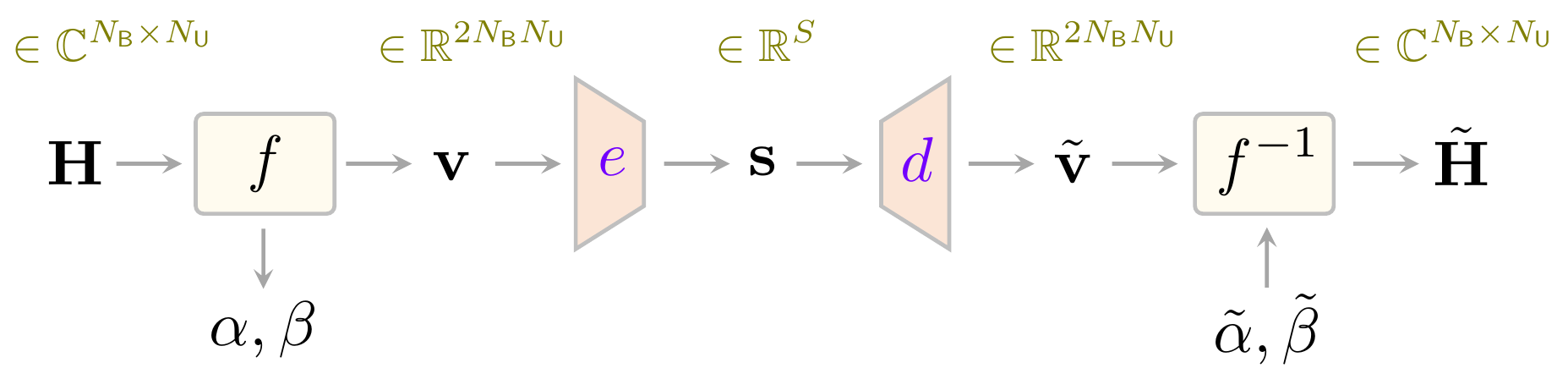}
  \vspace{-1em}
  \caption{Diagram of the designed autoencoder.}
  \label{fig_autoencoder}
\end{figure}

The first step in our method is to find a low-dimensional representation of the channel while preserving its temporal correlation property. This can be realized by training an autoencoder with an augmented loss function that encourages smooth time-varying features. The autoencoder consists of an encoder and a decoder; the encoder is trained to map $\Hm$ to a low-dimensional latent state $\sv$, while the decoder is trained to reconstruct $\Hm$ from the latent representation with minimal information loss. The architecture of the designed autoencoder is illustrated in Fig.~\ref{fig_autoencoder} and is detailed as follows.

\subsection{Data Preprocessing and Postprocessing} 
To adapt the complex-valued channel matrix for use with real-valued neural networks, we apply a preprocessing step that transforms the complex channel matrix $\Hm$ into a real-valued vector $\vv$ by extracting and concatenating its amplitude and phase components. Since these amplitudes and phases are typically on vastly different scales, the preprocessing function $f(\cdot)$ separates these components while normalizing them.

Let $\Hm_\mathsf{a}$ and $\Hm_\mathsf{p}$ denote the amplitude and phase matrices of $\Hm$, respectively, such that $\Hm = \Hm_\mathsf{a} \odot e^{j \Hm_\mathsf{p}}$, where $\odot$ denotes element-wise multiplication. The preprocessing function $f(\cdot)$ is defined as:
\begin{equation}
    \{\vv,\alpha,\beta\} = f(\Hm),
\end{equation}
where $\alpha = \mathrm{mean}(\Hm_\mathsf{a}),\quad \beta = \mathrm{std}(\Hm_\mathsf{a})$,
$\vv = [\big(\mathrm{vec}(\Hm_\mathsf{a})^\TT - \alpha\big)/\beta,\ \mathrm{vec}(\Hm_\mathsf{p})^\TT/\pi]^\TT$. Here, $\mathrm{mean}(\cdot)$ and $\mathrm{std}(\cdot)$ compute the mean and standard deviation of all elements in the input matrix, respectively. 

Conversely, the postprocessing function $f^{-1}(\cdot)$ reconstructs the channel matrix from the normalized amplitude and phase vectors as follows:
\begin{equation}\label{eq:post}
    \Hm = f^{-1}(\vv,\alpha,\beta),    
\end{equation}
where $\Hm = \Hm_\mathsf{a} \odot e^{j \Hm_\mathsf{p}}$ with
\begin{align}
    \Hm_\mathsf{a} &= \beta \cdot \mathrm{ivec}\big(\vv_{1:N_\Bt N_\Ut}, N_\Bt, N_\Ut\big) + \alpha,\notag\\
    \Hm_\mathsf{p} &= \pi \cdot \mathrm{ivec}\big(\vv_{N_\Bt N_\Ut+1:2N_\Bt N_\Ut}, N_\Bt, N_\Ut\big). \notag   
\end{align}
Here, $\mathrm{ivec}(\cdot, N_\Bt, N_\Ut)$ reshapes the input vector into a matrix with $N_\Bt$ rows and $N_\Ut$ columns.    

\subsection{Autoencoder Training}

The encoder $e(\cdot)$ and decoder $d(\cdot)$ are both \acp{MLP}, each consisting of multiple layers with trainable weights, biases, and activation functions. While autoencoders are widely used across various applications, the critical aspect of our approach lies in the design of the latent space. Specifically, the formulation of the training objective is paramount to achieving the desired latent representation properties.

For training, we utilize a static dataset collected from a fixed \ac{BS} and a set of users located at various positions, denoted as $\mathcal{H}=\{\Hm^{(k)},\pv^{(k)}\}_{k=1}^{K}$, where each $\Hm^{(k)}$ represents a channel matrix sample and $\pv^{(k)}$ denotes the corresponding user location. In this paper, we focus on users that do not move rapidly, so their movement distances are limited within a short timespan. Consequently, the temporal correlation of the channel features can be effectively characterized through the spatial correlation of the channels in $\mathcal{H}$.  Our autoencoder training should achieve two objectives: (i) channel information preservation and (ii) temporal correlation preservation. In the following, we elaborate on these two objectives and present the corresponding training methodology.

\subsubsection{Channel Information Preservation}
The first objective is to ensure that the autoencoder can accurately reconstruct the channel from the compressed latent state. The loss function for this objective is defined as follows:
\begin{equation}
    \mathcal{L}_{\mathsf{CI}} = \frac{1}{K}\sum_{k=1}^K \Big\| f\big(\Hm^{(k)}\big) -d\Big(e\big(f\big(\Hm^{(k)}\big)\big)+\nv\Big) \Big\|_2^2,    
\end{equation}
where $\nv$ is a zero-mean perturbation to avoid overfitting.     

\subsubsection{Temporal Correlation Preservation}
The second objective is to ensure that the learned latent representations possess the desired temporal correlation property. Specifically, for a slowly moving user, we require that its latent states vary slowly and smoothly over time. This critical property can be achieved by adopting a method similar to \emph{channel charting}~\cite{Studer2018Channel}. 

Channel charting is a method that learns a low-dimensional representation of wireless channels that preserves the spatial geometry among users. It trains a deep neural network that maps nearby users in physical space (e.g., $\pv^{(i)},\pv^{(j)}$) to similar latent states (e.g., $\zv^{(i)},\zv^{(j)}$). This builds a latent geometric manifold that reflects the underlying spatial topology of the radio environment by enforcing $\delta(\zv^{(i)},\zv^{(j)}) \approx \delta(\pv^{(i)},\pv^{(j)}),\ \forall i,j,$ where $\delta(\cdot,\cdot)$ represents a dissimilarity measure. Channel charting typically operates in a fully unsupervised manner without requiring location labels, exploring the local geometry through a feature extraction step that distills useful information from the channel into a feature geometry~\cite{Studer2018Channel}. In our context, we leverage ground-truth user position information to facilitate the training process, as our channel estimation problem requires both compression and accurate channel recovery from the latent state, which is more challenging than standard channel charting applications.

As mentioned, we leverage the spatial proximity of users to learn temporally correlated latent representations. Specifically, we design the second loss function similar to the channel charting loss as follows:
\begin{equation}\label{eq:LTC}
    \mathcal{L}_{\mathsf{TC}} = \|\Dm - \Bm \|_\mathsf{F}^2,    
\end{equation}
where $\Dm = \big(\bar{\Dm}-\mathrm{mean}(\bar{\Dm})\big)/\mathrm{std}(\bar{\Dm})$ and $\Bm = \big(\bar{\Bm}-\mathrm{mean}(\bar{\Bm})\big)/\mathrm{std}(\bar{\Bm})$ with $\bar{\Dm}, \bar{\Bm} \in\mathbb{R}^{K\times K}$ defined as
\begin{align}
    [\bar{\Dm}]_{i,j} &= \big\|e\big(f(\Hm^{(i)})\big) - e\big(f(\Hm^{(j)})\big)\big\|_2^2,\\
    [\bar{\Bm}]_{i,j} &= \big\| \pv^{(i)} - \pv^{(j)} \big\|_2^2.
\end{align} 

\subsubsection{Overall Training Objective}
The overall training objective of the autoencoder combines the two loss functions, resulting in the following optimization problem:
\begin{equation}
    \min_{e(\cdot),\, d(\cdot)}\ \mathcal{L}_{\mathsf{CI}} + \lambda \mathcal{L}_{\mathsf{TC}},
\end{equation}
where $\lambda$ is a hyperparameter that controls the trade-off between the two objectives.

\section{Step 2: Latent Dynamic Tracking}\label{sec:LSTM}

Having trained the autoencoder, we can now leverage it to track the latent state across multiple channel coherence intervals using limited pilot observations. We suggest employing \ac{LSTM} networks~\cite{Hochreiter1997Long} for this tracking task, as they excel at handling temporal sequences and capturing short-term and long-term dependencies in the evolving observations.

\subsection{LSTM Network Design}

\begin{figure}[t]
  \centering
  \includegraphics[width=\linewidth]{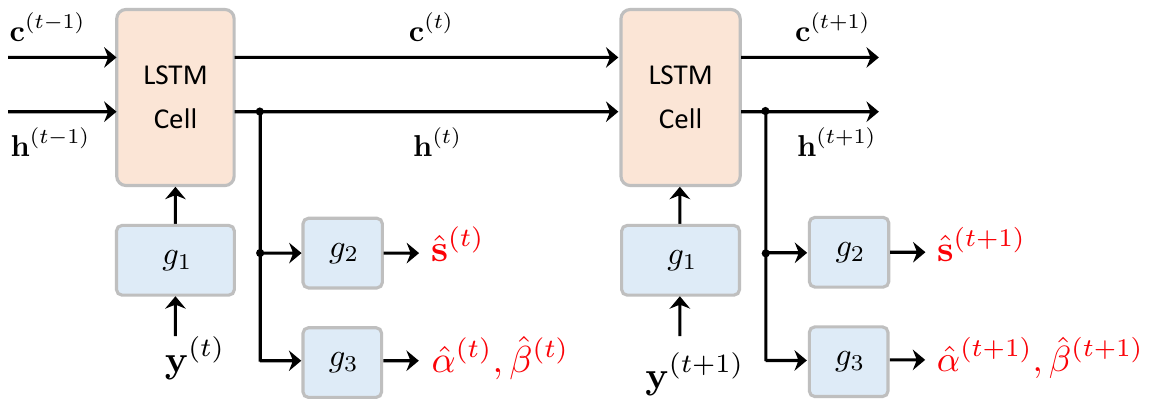}
  \vspace{-1em}
  \caption{Diagram of the designed LSTM network for latent state tracking across channel coherence intervals.}
  \label{fig_LSTM}
\end{figure} 

The architecture of the designed \ac{LSTM} network is illustrated in Fig.~\ref{fig_LSTM}. Consider a sequence of $T$ consecutive channel coherence intervals. We denote the received pilot signals at the \ac{BS} during the $t^{\text{th}}$ interval as $\Ym^{(t)}$, for $t=1,2,\dots,T$, whose expression follows~\eqref{eq:SigModel}. We reshape it as $\yv^{(t)}=[\mathrm{real}(\mathrm{vec}(\Ym^{(t)}))^\TT,\mathrm{imag}(\mathrm{vec}(\Ym^{(t)}))^\TT]^\TT$, where $\mathrm{real}(\cdot)$ and $\mathrm{imag}(\cdot)$ extract the real and imaginary parts of a complex vector, respectively. Then, we input the sequence of received pilot signals $\{\yv^{(t)}\}_{t=1}^T$ into the \ac{LSTM} network through a \ac{MLP} $g_1(\cdot)$. The \ac{LSTM} cell state $\cv^{(t)}$ and hidden state $\hv^{(t)}$ are initialized to zero, and updated with each time step based on the input and previous states. Next, we output the latent states $\{\hat{\sv}^{(t)}\}_{t=1}^T$ and normalization scalars $\{\hat{\alpha}^{(t)},\hat{\beta}^{(t)}\}_{t=1}^T$ through \acp{MLP} $g_2(\cdot)$ and $g_3(\cdot)$, respectively, from the hidden states of the \ac{LSTM} network. 

\subsection{LSTM Training}
During the training of the \ac{LSTM} network, the decoder weights are pretrained and fixed. We optimize only the weights of the \ac{LSTM} network and the associated \acp{MLP} $g_1$, $g_2$, and $g_3$ to minimize the estimation error of $\{{\sv}^{(t)},{\alpha}^{(t)},{\beta}^{(t)}\}_{t=1}^T$. The ground-truth values for these latent states and normalization scalars are obtained by generating a set of $T$-length channel sequences following continuous trajectories from the dataset $\mathcal{H}$ and passing these true channel matrices through the preprocessing function and trained encoder. The \ac{LSTM} network inputs are reshaped received signal sequences obtained from~\eqref{eq:SigModel} based on these ground-truth channel matrices. The loss function for training the \ac{LSTM} network is defined as
\begin{multline}
    \mathcal{L}_{\mathsf{LSTM}} = \frac{1}{T}\sum_{t=1}^T \Big(\big\|\sv^{(t)} - \hat{\sv}^{(t)}\big\|_2^2 + \lambda_\alpha\big|{\alpha}^{(t)} - \hat{\alpha}^{(t)}\big|_2^2 \\
    + \lambda_\beta \big|{\beta}^{(t)} - \hat{\beta}^{(t)}\big|^2\Big),
\end{multline}
where $\lambda_\alpha$ and $\lambda_\beta$ are hyperparameters that control the trade-off between the three loss terms.

\subsection{LSTM Inference}

In the inference phase, the received pilot signals at each coherence interval, $\yv^{(t)}$, is input into the trained \ac{LSTM} network to estimate the latent states $\hat{\sv}^{(t)}$ and normalization scalars $\hat{\alpha}^{(t)},\hat{\beta}^{(t)}$. Subsequently, we can reconstruct the full channel matrices $\{\hat{\Hm}^{(t)}\}_{t=1}^T$ using the postprocessing function and trained decoder (according to~\eqref{eq:post}) as
\begin{equation}\label{eq:Hhat}
    \hat{\Hm}^{(t)} \!=\! f^{-1}\Big(d\big(\hat{\sv}^{(t)}\big), \hat{\alpha}^{(t)}, \hat{\beta}^{(t)}\Big),\ t=1,2,\dots,T. 
\end{equation} 

\subsection{Additional Considerations}

A key consideration in designing the \ac{LSTM} network to track the latent state instead of the full channel matrix is that the latent state dimension $S$ is significantly smaller than the total number of channel coefficients. More importantly, the scale of the designed \ac{LSTM} network shown in Fig.~\ref{fig_LSTM} is independent of the number of antennas at both the \ac{BS} and \ac{UE}. This property is crucial for practical deployment in massive \ac{MIMO} systems. 

While large-scale network training is not entirely eliminated, as it remains required during the autoencoder training phase, this separation provides significant advantages. The static autoencoder training is inherently simpler than sequential \ac{LSTM} training, since the former requires only individual channel samples, whereas the latter demands temporally ordered sequences of channels.

In this paper, we fix the pilot symbols $\Sm$ and the precoding and combining matrices $\Fm$ and $\Wm$ during both the training and inference phases to maintain simplicity and focus on the core methodology. However, these configurations can be jointly optimized alongside the deep network to further enhance estimation performance~\cite{Sohrabi2022Active}, which will be explored in future research.

\section{Numerical Results}

\begin{figure}[t]
    \centering
    % ----- a -----
    \begin{subfigure}{0.75\linewidth}
        \centering
        \includegraphics[width=\linewidth]{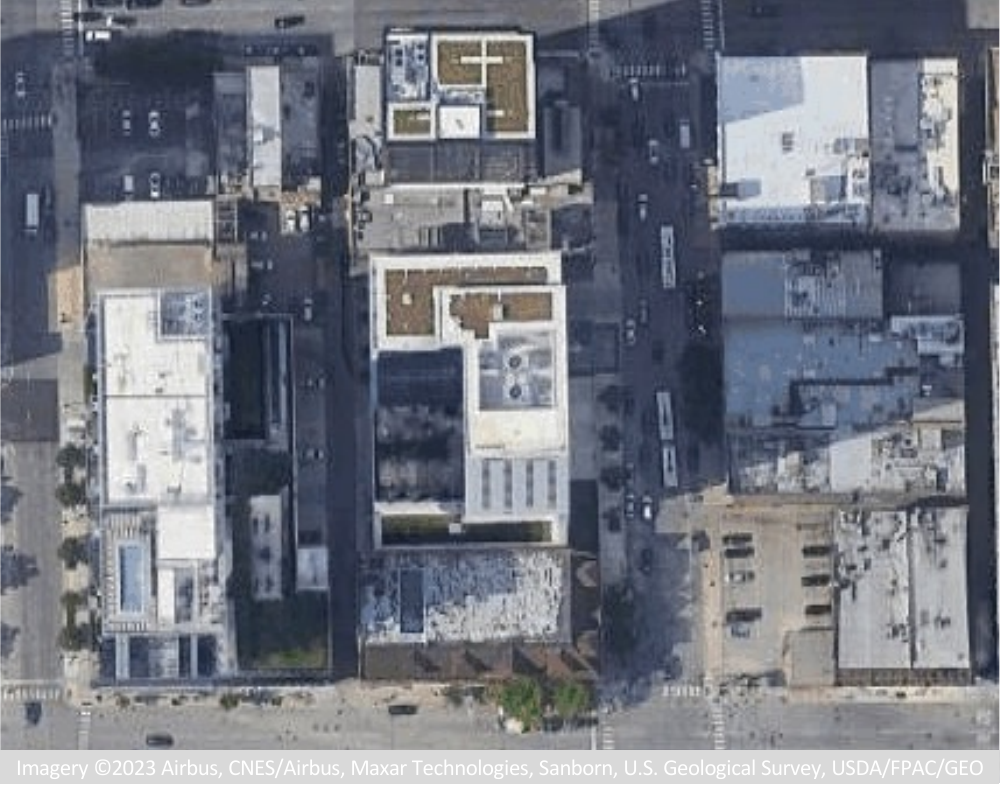}  
        \caption{Photograph of the Simulated Environment }
        \label{fig:pt}
    \end{subfigure}
    
    \vspace{0.3cm}

    % ----- b -----
    \begin{subfigure}{\linewidth}
        \centering
        \includegraphics[width=\linewidth]{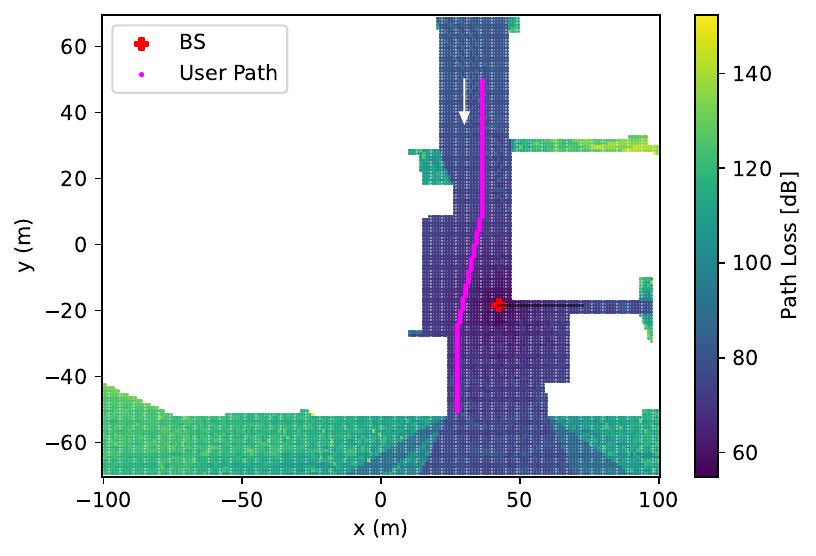}  
        \caption{Path Loss Heatmap with User Path Overlay}
        \label{fig:path}
    \end{subfigure}

    \caption{Evaluation Scenario from DeepMIMO Dataset~\cite{Alkhateeb2019}}
    \label{fig:deepmimo}
\end{figure}

We train the proposed deep models using the DeepMIMO dataset~\cite{Alkhateeb2019} with $1111$ channel samples from the Chicago city scenario. The \ac{BS} is equipped with a $10 \times 10 = 100$ antenna \ac{UPA} ($N_\Bt = 100$), while the \ac{UE} has a $2\times 2 = 4$ antenna \ac{UPA} ($N_\Ut = 4$). The carrier frequency is $3.5$~GHz, and the system bandwidth is $10$~MHz divided into $512$ subcarriers. The transmit power is $45$~dBm, and the noise power is $-95$~dBm. The signaling overhead is set to $M_\Bt M_\Ut = 96$, much less than the $N_\Bt N_\Ut = 400$ unknown channel coefficients. The latent state dimension is $S=64$. The encoder $e(\cdot)$ consists of 2 hidden layers with widths [1280, 256], while the decoder $d(\cdot)$ has 2 hidden layers with widths [256, 1280]. The \ac{LSTM} network contains 3 layers with 64 hidden units. All \acp{MLP} use ReLU activation for hidden layers and linear activation for output layers. Both networks are implemented in PyTorch using the Adam optimizer and evaluated on a user trajectory illustrated in Fig.~\ref{fig:deepmimo}, with channel and received signal data generated using ray-tracing simulations from a real-world environment~\cite{Alkhateeb2019}.

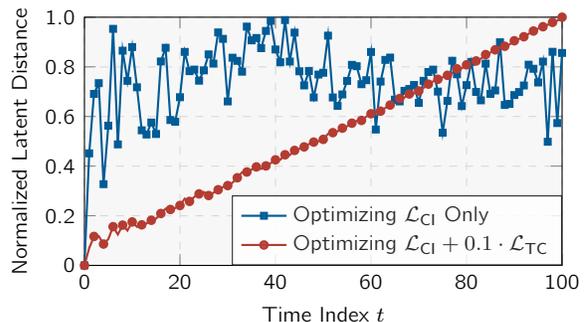
\begin{figure}[t]
    \centering
    % ---------------------------
% Local Font Settings (only inside this group)
% ---------------------------
\renewcommand{\sfdefault}{cmbr} % Example: TeX Gyre Heros (sans-serif)
\newcommand{\ttk}[1]{\footnotesize\textsf{#1}} % Tick helper

% ---------------------------
% PGFPlots Local Styles
% ---------------------------
\pgfplotsset{
    every axis/.append style={
        label style={font=\sffamily\footnotesize},      % Axis labels font
        legend style={font=\sffamily\footnotesize},     % Legend font
        tick label style={font=\sffamily\footnotesize}  % Tick labels fallback
    }
}

% ---------------------------
% TikZ Plot: Normalized Single Y-Axis
% ---------------------------
\begin{tikzpicture}
\begin{axis}[%
    width=2.5in, height=1.3in, scale only axis,
    xmin=0, xmax=100, ymin=0, ymax=1,
    xlabel={Time Index $t$}, 
    ylabel={Normalized Latent Distance},
    xtick={0,20,40,60,80,100},
    ytick={0,0.2,0.4,0.6,0.8,1.0},
    xticklabels={\ttk{0},\ttk{20},\ttk{40},\ttk{60},\ttk{80},\ttk{100}},
    yticklabels={\ttk{0},\ttk{0.2},\ttk{0.4},\ttk{0.6},\ttk{0.8},\ttk{1.0}},
    axis background/.style={fill=black!3},
    xmajorgrids=true, 
    ymajorgrids=true,
    grid style={dashed, gray!30},
    legend style={at={(1,0)}, anchor=south east, legend cell align=left, align=left, draw=white!15!black, font=\sffamily\footnotesize, fill opacity=0.85}
]

% ===========================
% Curve 1: Optimizing L_CI Only (Red)
% ===========================
\addplot[
    line width=0.8pt,
    color=profBlue,
    mark=square*,
    mark size=1.1pt
]
table[row sep=crcr]{%
0   0.0000\\
1   0.4514\\
2   0.6911\\
3   0.7343\\
4   0.3273\\
5   0.5630\\
6   0.9531\\
7   0.4876\\
8   0.8652\\
9   0.7437\\
10  0.8797\\
11  0.7177\\
12  0.5443\\
13  0.5274\\
14  0.5757\\
15  0.5296\\
16  0.8217\\
17  0.8768\\
18  0.5859\\
19  0.5788\\
20  0.6777\\
21  0.8602\\
22  0.7826\\
23  0.7889\\
24  0.7443\\
25  0.7858\\
26  0.8503\\
27  0.8131\\
28  0.9387\\
29  0.9128\\
30  0.6607\\
31  0.8368\\
32  0.8241\\
33  0.7805\\
34  0.9621\\
35  0.9067\\
36  0.9162\\
37  0.8753\\
38  0.9447\\
39  0.9855\\
40  0.8700\\
41  0.8186\\
42  0.9885\\
43  0.8214\\
44  0.9380\\
45  0.7821\\
46  0.7252\\
47  0.7831\\
48  0.6768\\
49  0.7690\\
50  0.7762\\
51  0.9257\\
52  0.6766\\
53  0.6428\\
54  0.6889\\
55  0.7433\\
56  0.8080\\
57  0.8027\\
58  0.7301\\
59  0.7462\\
60  0.8599\\
61  0.5474\\
62  0.7408\\
63  0.8280\\
64  0.8374\\
65  0.6679\\
66  0.6607\\
67  0.7034\\
68  0.7108\\
69  0.7269\\
70  0.6547\\
71  0.7437\\
72  0.7793\\
73  0.7888\\
74  0.7004\\
75  0.5346\\
76  0.6627\\
77  0.8245\\
78  0.7700\\
79  0.6420\\
80  0.7264\\
81  0.8199\\
82  0.6989\\
83  0.6652\\
84  0.8128\\
85  0.6910\\
86  0.7050\\
87  0.8991\\
88  0.6465\\
89  0.6508\\
90  0.6863\\
91  0.6986\\
92  0.7249\\
93  0.8088\\
94  0.7923\\
95  0.7370\\
96  0.8212\\
97  0.4982\\
98  0.8606\\
99  0.5735\\
100 0.8560\\
};
\addlegendentry{Optimizing $\mathcal{L}_{\mathsf{CI}}$ Only}

% ===========================
% Curve 2: Optimizing L_CI + λ L_TC (Blue)
% ===========================
\addplot[
    line width=0.8pt,
    color=profRed,
    mark=*,
    mark repeat=2,
    mark size=1.4pt
]
table[row sep=crcr]{%
0   0.000000\\
1   0.0699\\
2   0.1166\\
3   0.1179\\
4   0.0860\\
5   0.1060\\
6   0.1560\\
7   0.1245\\
8   0.1623\\
9   0.1403\\
10  0.1751\\
11  0.1552\\
12  0.1634\\
13  0.1690\\
14  0.1818\\
15  0.1887\\
16  0.2091\\
17  0.2288\\
18  0.2253\\
19  0.2309\\
20  0.2414\\
21  0.2690\\
22  0.2579\\
23  0.2759\\
24  0.2881\\
25  0.2959\\
26  0.2815\\
27  0.2926\\
28  0.3049\\
29  0.3157\\
30  0.3216\\
31  0.3333\\
32  0.3531\\
33  0.3751\\
34  0.3763\\
35  0.3860\\
36  0.3968\\
37  0.3885\\
38  0.4009\\
39  0.4119\\
40  0.4254\\
41  0.4257\\
42  0.4458\\
43  0.4445\\
44  0.4633\\
45  0.4693\\
46  0.4775\\
47  0.4888\\
48  0.4968\\
49  0.5002\\
50  0.5078\\
51  0.5300\\
52  0.5348\\
53  0.5401\\
54  0.5530\\
55  0.5605\\
56  0.5730\\
57  0.5823\\
58  0.5841\\
59  0.5994\\
60  0.6110\\
61  0.6188\\
62  0.6214\\
63  0.6345\\
64  0.6460\\
65  0.6543\\
66  0.6724\\
67  0.6763\\
68  0.6907\\
69  0.6926\\
70  0.7029\\
71  0.7148\\
72  0.7304\\
73  0.7413\\
74  0.7499\\
75  0.7577\\
76  0.7659\\
77  0.7740\\
78  0.7919\\
79  0.7938\\
80  0.8082\\
81  0.8174\\
82  0.8238\\
83  0.8330\\
84  0.8494\\
85  0.8574\\
86  0.8686\\
87  0.8800\\
88  0.8830\\
89  0.8902\\
90  0.9051\\
91  0.9126\\
92  0.9214\\
93  0.9330\\
94  0.9444\\
95  0.9543\\
96  0.9617\\
97  0.9694\\
98  0.9792\\
99  0.9858\\
100  1.0000\\
};
\addlegendentry{Optimizing $\mathcal{L}_{\mathsf{CI}} + 0.1 \cdot \mathcal{L}_{\mathsf{TC}}$}

\end{axis}
\end{tikzpicture}
    \vspace{-2.5em}
    \caption{Comparison of normalized latent distance for two autoencoder training strategies: (i) without temporal correlation loss (blue square) and (ii) with temporal correlation loss (red circle). The latent distance is calculated as $\|\sv^{(t)}-\sv^{(0)}\|_2$ and normalized for visualization.}
    \label{fig_s}
\end{figure}

A critical aspect of the proposed method is the design of the latent space to preserve temporal correlation, achieved through the inclusion of the loss term $\mathcal{L}_\mathsf{TC}$ in~\eqref{eq:LTC}. To demonstrate the impact of this design, Fig.~\ref{fig_s} compares the temporal evolution of latent state distances under two training scenarios. Without the temporal correlation loss (blue square), the learned latent representation exhibits no discernible temporal structure, as the latent distance to the initial time step $\|\sv^{(t)}-\sv^{(0)}\|_2$ fluctuates erratically over time. This significantly complicates the tracking task for the \ac{LSTM} network. In contrast, when the temporal correlation loss is incorporated during autoencoder training (red circles), the latent states evolve smoothly and gradually over time, creating a more tractable tracking problem that enables the \ac{LSTM} network to effectively capture the underlying dynamics (with only negligible reconstruction accuracy loss).

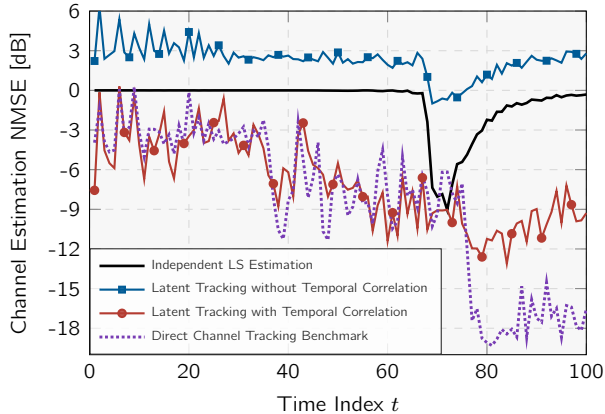
\begin{figure}[t]
    \centering
    % ---------------------------
% Local Font Settings (only inside this group)
% ---------------------------
\renewcommand{\sfdefault}{cmbr} % Example: TeX Gyre Heros (sans-serif)
\newcommand{\ttk}[1]{\footnotesize\textsf{#1}} % Tick helper

% ---------------------------
% PGFPlots Local Styles
% ---------------------------
\pgfplotsset{
    every axis/.append style={
        label style={font=\sffamily\footnotesize},      % Axis labels font
        legend style={font=\sffamily\footnotesize},     % Legend font
        tick label style={font=\sffamily\footnotesize}  % Tick labels fallback
    }
}

% ---------------------------
% TikZ Plot
% ---------------------------
\begin{tikzpicture}
\begin{axis}[%
    % --- Plot Size ---
    width=2.6in, height=1.8in, scale only axis,
    % --- Axis Range ---
    xmin=0, xmax=100, ymin=-20, ymax=6,
    % --- Axis Labels ---
    xlabel={Time Index $t$}, 
    ylabel={Channel Estimation NMSE [dB]},
    % --- Custom Tick Positions ---
    xtick={0,20,40,60,80,100},
    ytick={-18,-15,-12,-9,-6,-3,0,3,6},
    % --- Tick Labels using \ttk ---
    xticklabels={\ttk{0},\ttk{20},\ttk{40},\ttk{60},\ttk{80},\ttk{100}},
    yticklabels={\ttk{-18},\ttk{-15},\ttk{-12},\ttk{-9},\ttk{-6},\ttk{-3},\ttk{0},\ttk{3},\ttk{6}},
    % --- Grid & Background ---
    axis background/.style={fill=black!3},
    xmajorgrids=true, 
    ymajorgrids=true,
    grid style={dashed, gray!30},
    % --- Legend setup ---
    legend style={at={(0,0)}, anchor=south west, legend cell align=left, align=left, draw=white!15!black, font=\sffamily\tiny, fill opacity=0.85}
]

% ===========================
% Curve 1: MC-unaware OMP
% ===========================
\addplot[
    line width=1pt,
    color=black,
]table[row sep=crcr]{%
1	-0.000617\\
2	-0.002291\\
3	-0.000757\\
4	-0.000609\\
5	-0.000822\\
6	-0.002211\\
7	-0.000806\\
8	-0.000706\\
9	-0.003604\\
10	-0.000958\\
11	-0.000827\\
12	-0.003584\\
13	-0.001106\\
14	-0.000889\\
15	-0.004287\\
16	-0.001324\\
17	-0.001952\\
18	-0.002008\\
19	-0.001276\\
20	-0.004069\\
21	-0.001282\\
22	-0.006758\\
23	-0.001281\\
24	-0.004492\\
25	-0.002017\\
26	-0.002842\\
27	-0.001345\\
28	-0.001653\\
29	-0.001722\\
30	-0.001685\\
31	-0.001834\\
32	-0.001910\\
33	-0.002203\\
34	-0.002218\\
35	-0.002104\\
36	-0.002217\\
37	-0.002313\\
38	-0.002222\\
39	-0.002307\\
40	-0.002279\\
41	-0.002213\\
42	-0.002805\\
43	-0.003308\\
44	-0.005063\\
45	-0.005801\\
46	-0.007346\\
47	-0.012453\\
48	-0.015016\\
49	-0.018296\\
50	-0.020561\\
51	-0.026017\\
52	-0.028353\\
53	-0.024943\\
54	-0.019379\\
55	-0.004097\\
56	-0.003493\\
57	-0.013727\\
58	-0.044368\\
59	-0.055508\\
60	-0.021698\\
61	-0.011315\\
62	-0.065912\\
63	-0.084283\\
64	-0.023614\\
65	-0.205278\\
66	-0.234996\\
67	-0.212312\\
68	-2.387584\\
69	-7.367558\\
70	-8.041702\\
71	-7.857262\\
72	-9.044137\\
73	-7.131414\\
74	-5.578469\\
75	-5.419632\\
76	-4.861221\\
77	-4.072104\\
78	-3.333052\\
79	-3.040074\\
80	-2.224304\\
81	-2.291733\\
82	-1.623844\\
83	-1.618294\\
84	-1.144717\\
85	-1.218503\\
86	-0.970388\\
87	-1.081243\\
88	-0.955420\\
89	-0.698751\\
90	-0.733436\\
91	-0.770896\\
92	-0.561022\\
93	-0.578128\\
94	-0.363346\\
95	-0.520711\\
96	-0.388737\\
97	-0.342162\\
98	-0.400586\\
99	-0.365743\\
100	-0.314067\\
};
\addlegendentry{Independent LS Estimation}

\addplot[
    line width=0.8pt,
    color=profBlue,
    mark=square*,
    mark repeat=6, 
    mark size=1.1pt
]table[row sep=crcr]{%
1	2.222193\\
2	6.298187\\
3	2.384946\\
4	2.681467\\
5	2.967251\\
6	5.383020\\
7	2.869824\\
8	2.500989\\
9	4.298987\\
10	3.213342\\
11	2.358105\\
12	5.013164\\
13	2.990715\\
14	2.755288\\
15	4.517180\\
16	2.786416\\
17	3.572290\\
18	3.256056\\
19	2.262346\\
20	4.416265\\
21	2.493468\\
22	4.325209\\
23	2.468967\\
24	3.403210\\
25	2.603723\\
26	3.409704\\
27	2.599640\\
28	2.023127\\
29	2.854516\\
30	2.476099\\
31	2.580277\\
32	2.320104\\
33	2.507055\\
34	2.417028\\
35	2.283957\\
36	2.650613\\
37	2.509825\\
38	2.685132\\
39	2.550866\\
40	2.632367\\
41	2.367607\\
42	2.347692\\
43	2.205905\\
44	2.491459\\
45	2.329313\\
46	2.200996\\
47	2.899381\\
48	2.467519\\
49	2.018049\\
50	2.870548\\
51	2.095657\\
52	2.362618\\
53	2.021388\\
54	2.145819\\
55	2.128940\\
56	2.509503\\
57	2.432407\\
58	2.722808\\
59	2.307526\\
60	1.933200\\
61	1.701711\\
62	2.233210\\
63	1.987430\\
64	2.127191\\
65	1.889880\\
66	2.836223\\
67	2.394774\\
68	1.016492\\
69	-0.994370\\
70	-0.827079\\
71	-0.616678\\
72	-0.704736\\
73	-0.178153\\
74	-0.532882\\
75	-0.227083\\
76	0.093942\\
77	0.322313\\
78	1.051241\\
79	0.611949\\
80	1.192176\\
81	0.861090\\
82	1.932140\\
83	1.267072\\
84	2.056851\\
85	1.873208\\
86	2.076561\\
87	1.707277\\
88	2.166261\\
89	2.612975\\
90	2.317031\\
91	2.238459\\
92	2.242282\\
93	2.422913\\
94	2.265106\\
95	2.087184\\
96	2.939907\\
97	2.913488\\
98	2.764760\\
99	2.380131\\
100	2.796547\\
};
\addlegendentry{Latent Tracking without Temporal Correlation}

\addplot[
    line width=0.8pt,
    color=profRed,
    mark=*,
    mark repeat=6,
    mark size=1.4pt
]table[row sep=crcr]{%
1	-7.563859\\
2	-0.144552\\
3	-4.544632\\
4	-5.509899\\
5	-5.885967\\
6	0.298622\\
7	-3.181615\\
8	-3.776710\\
9	-0.418470\\
10	-4.118938\\
11	-6.721049\\
12	-3.505355\\
13	-4.561488\\
14	-3.355937\\
15	-1.601841\\
16	-5.394789\\
17	-6.665487\\
18	-3.823839\\
19	-4.019122\\
20	-3.464281\\
21	-3.008473\\
22	-1.363634\\
23	-3.987235\\
24	-3.841170\\
25	-2.441001\\
26	-2.315788\\
27	-0.620762\\
28	-3.799357\\
29	-3.257963\\
30	-4.734822\\
31	-4.162351\\
32	-4.912052\\
33	-2.601749\\
34	-3.350061\\
35	-4.378759\\
36	-6.429743\\
37	-7.061470\\
38	-8.779309\\
39	-5.897053\\
40	-6.193592\\
41	-8.035561\\
42	-3.328270\\
43	-2.466567\\
44	-4.575472\\
45	-5.629972\\
46	-5.890557\\
47	-6.376903\\
48	-8.037706\\
49	-7.112944\\
50	-5.514368\\
51	-7.114044\\
52	-7.517138\\
53	-7.378724\\
54	-6.791550\\
55	-8.042958\\
56	-8.035018\\
57	-10.629262\\
58	-8.695235\\
59	-7.730317\\
60	-11.401050\\
61	-9.258787\\
62	-11.000979\\
63	-6.627837\\
64	-9.342094\\
65	-7.638269\\
66	-10.445928\\
67	-6.611495\\
68	-10.539593\\
69	-9.301639\\
70	-9.074835\\
71	-9.089652\\
72	-9.039104\\
73	-10.000270\\
74	-7.256028\\
75	-10.091014\\
76	-10.896651\\
77	-11.955518\\
78	-11.964598\\
79	-12.604634\\
80	-11.236645\\
81	-11.355686\\
82	-10.617335\\
83	-10.833831\\
84	-13.111937\\
85	-10.848103\\
86	-10.074017\\
87	-9.879788\\
88	-11.461500\\
89	-9.133128\\
90	-8.979375\\
91	-11.171991\\
92	-10.148769\\
93	-9.197551\\
94	-7.613640\\
95	-10.018095\\
96	-7.199815\\
97	-8.650016\\
98	-9.908057\\
99	-9.837702\\
100	-9.284947\\
};
\addlegendentry{Latent Tracking with Temporal Correlation}

\addplot[
    line width=1.1pt,
    densely dotted,
    color=profPurple
]table[row sep=crcr]{%
1	-3.959315\\
2	-1.348294\\
3	-3.878840\\
4	-4.660446\\
5	-4.516737\\
6	0.062309\\
7	-3.388763\\
8	-2.774313\\
9	0.241435\\
10	-3.199618\\
11	-5.019270\\
12	-2.930508\\
13	-2.944313\\
14	-3.133798\\
15	-4.052627\\
16	-3.358131\\
17	-4.791818\\
18	-2.673688\\
19	-4.071506\\
20	-0.157166\\
21	-3.212180\\
22	-2.063469\\
23	-3.783963\\
24	-2.511290\\
25	-3.681763\\
26	-3.065961\\
27	-4.066188\\
28	-3.862576\\
29	-3.638048\\
30	-3.273950\\
31	-2.761737\\
32	-4.255429\\
33	-3.066070\\
34	-3.540043\\
35	-2.234526\\
36	-4.288840\\
37	-8.139887\\
38	-10.572159\\
39	-11.248397\\
40	-8.321854\\
41	-7.588325\\
42	-2.396696\\
43	-5.248928\\
44	-8.123966\\
45	-6.621750\\
46	-3.948334\\
47	-10.668119\\
48	-11.320151\\
49	-6.954201\\
50	-6.364966\\
51	-7.149905\\
52	-9.465332\\
53	-8.810968\\
54	-6.754415\\
55	-8.539641\\
56	-8.481366\\
57	-10.088296\\
58	-6.135889\\
59	-4.496769\\
60	-7.986580\\
61	-10.450658\\
62	-9.905854\\
63	-5.041632\\
64	-7.964253\\
65	-7.523311\\
66	-8.864819\\
67	-6.455309\\
68	-7.695484\\
69	-10.274229\\
70	-9.371352\\
71	-6.190892\\
72	-6.175210\\
73	-7.767512\\
74	-8.537618\\
75	-7.042136\\
76	-11.987805\\
77	-17.423570\\
78	-16.521394\\
79	-18.759010\\
80	-19.088722\\
81	-19.305691\\
82	-18.816731\\
83	-18.091515\\
84	-18.431303\\
85	-18.072386\\
86	-18.663878\\
87	-14.833785\\
88	-18.693385\\
89	-15.824921\\
90	-17.239788\\
91	-17.536302\\
92	-14.533445\\
93	-17.095023\\
94	-16.498546\\
95	-18.695158\\
96	-16.283035\\
97	-16.399300\\
98	-17.842603\\
99	-18.161158\\
100	-16.539144\\
};
\addlegendentry{Direct Channel Tracking Benchmark}

\end{axis}
\end{tikzpicture}
    \vspace{-2.5em}
    \caption{Channel Estimation NMSE Over Time for $10\times 10$ BS.}
    \label{fig_a100}
\end{figure}

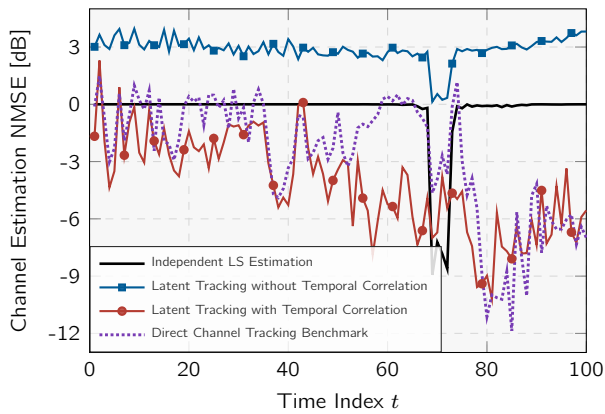
\begin{figure}[t]
    \centering
    % ---------------------------
% Local Font Settings (only inside this group)
% ---------------------------
\renewcommand{\sfdefault}{cmbr} % Example: TeX Gyre Heros (sans-serif)
\newcommand{\ttk}[1]{\footnotesize\textsf{#1}} % Tick helper

% ---------------------------
% PGFPlots Local Styles
% ---------------------------
\pgfplotsset{
    every axis/.append style={
        label style={font=\sffamily\footnotesize},      % Axis labels font
        legend style={font=\sffamily\footnotesize},     % Legend font
        tick label style={font=\sffamily\footnotesize}  % Tick labels fallback
    }
}

% ---------------------------
% TikZ Plot
% ---------------------------
\begin{tikzpicture}
\begin{axis}[%
    % --- Plot Size ---
    width=2.6in, height=1.8in, scale only axis,
    % --- Axis Range ---
    xmin=0, xmax=100, ymin=-13, ymax=5,
    % --- Axis Labels ---
    xlabel={Time Index $t$}, 
    ylabel={Channel Estimation NMSE [dB]},
    % --- Custom Tick Positions ---
    xtick={0,20,40,60,80,100},
    ytick={-12,-9,-6,-3,0,3},
    % --- Tick Labels using \ttk ---
    xticklabels={\ttk{0},\ttk{20},\ttk{40},\ttk{60},\ttk{80},\ttk{100}},
    yticklabels={\ttk{-12},\ttk{-9},\ttk{-6},\ttk{-3},\ttk{0},\ttk{3}},
    % --- Grid & Background ---
    axis background/.style={fill=black!3},
    xmajorgrids=true, 
    ymajorgrids=true,
    grid style={dashed, gray!30},
    % --- Legend setup ---
    legend style={at={(0,0)}, anchor=south west, legend cell align=left, align=left, draw=white!15!black, font=\sffamily\tiny, fill opacity=0.85}
]

% ===========================
% Curve 1: MC-unaware OMP
% ===========================
\addplot[
    line width=1pt,
    color=black,
]table[row sep=crcr]{%
1	-0.000039\\
2	-0.000158\\
3	-0.000034\\
4	-0.000024\\
5	-0.000065\\
6	-0.000101\\
7	-0.000031\\
8	-0.000042\\
9	-0.000250\\
10	-0.000050\\
11	-0.000043\\
12	-0.000267\\
13	-0.000061\\
14	-0.000076\\
15	-0.000192\\
16	-0.000076\\
17	-0.000216\\
18	-0.000095\\
19	-0.000119\\
20	-0.000179\\
21	-0.000124\\
22	-0.000314\\
23	-0.000104\\
24	-0.000239\\
25	-0.000183\\
26	-0.000224\\
27	-0.000048\\
28	-0.000044\\
29	-0.000036\\
30	-0.000048\\
31	-0.000066\\
32	-0.000089\\
33	-0.000092\\
34	-0.000125\\
35	-0.000128\\
36	-0.000168\\
37	-0.000206\\
38	-0.000234\\
39	-0.000268\\
40	-0.000307\\
41	-0.000352\\
42	-0.000439\\
43	-0.000712\\
44	-0.000859\\
45	-0.000954\\
46	-0.001040\\
47	-0.001023\\
48	-0.000976\\
49	-0.000746\\
50	-0.000455\\
51	-0.000358\\
52	-0.001092\\
53	-0.002111\\
54	-0.002770\\
55	-0.000175\\
56	-0.001410\\
57	-0.004647\\
58	-0.004140\\
59	-0.007497\\
60	-0.004834\\
61	-0.006306\\
62	-0.010217\\
63	-0.026972\\
64	-0.016373\\
65	-0.009702\\
66	-0.103151\\
67	-0.246433\\
68	-0.189945\\
69	-8.937861\\
70	-7.224475\\
71	-7.831209\\
72	-8.603885\\
73	-1.438791\\
74	-0.112133\\
75	-0.207695\\
76	-0.008204\\
77	-0.122256\\
78	-0.058515\\
79	-0.092827\\
80	-0.077797\\
81	-0.093534\\
82	-0.053078\\
83	-0.148068\\
84	-0.041644\\
85	-0.116674\\
86	-0.058941\\
87	-0.028843\\
88	-0.048375\\
89	-0.005680\\
90	-0.003247\\
91	-0.001849\\
92	-0.004135\\
93	-0.008257\\
94	-0.009731\\
95	-0.011431\\
96	-0.004391\\
97	-0.002734\\
98	-0.001741\\
99	-0.000904\\
100	-0.000920\\
};
\addlegendentry{Independent LS Estimation}

\addplot[
    line width=0.8pt,
    color=profBlue,
    mark=square*,
    mark repeat=6, 
    mark size=1.1pt
]table[row sep=crcr]{%
1	3.010970\\
2	3.643123\\
3	3.105086\\
4	2.895995\\
5	3.610352\\
6	3.946251\\
7	3.097804\\
8	3.219187\\
9	3.954395\\
10	3.120048\\
11	2.873611\\
12	3.317758\\
13	3.094536\\
14	3.001831\\
15	3.902338\\
16	2.759365\\
17	3.856740\\
18	3.065252\\
19	3.148924\\
20	3.442711\\
21	2.835694\\
22	3.501174\\
23	2.907952\\
24	3.092465\\
25	2.811197\\
26	2.948936\\
27	2.830492\\
28	3.188410\\
29	2.612222\\
30	3.059454\\
31	2.515774\\
32	3.025725\\
33	2.631413\\
34	2.783514\\
35	3.254640\\
36	2.623250\\
37	3.168088\\
38	2.738617\\
39	2.543645\\
40	3.411505\\
41	3.025198\\
42	2.363010\\
43	2.968850\\
44	2.751666\\
45	2.959281\\
46	2.615379\\
47	2.562928\\
48	2.545311\\
49	2.730403\\
50	2.645372\\
51	3.015541\\
52	2.848482\\
53	2.756628\\
54	2.438753\\
55	2.657625\\
56	2.705269\\
57	2.499814\\
58	2.514940\\
59	2.308270\\
60	2.736850\\
61	2.966607\\
62	2.684545\\
63	2.817932\\
64	2.778264\\
65	2.512157\\
66	2.654712\\
67	2.452255\\
68	2.754607\\
69	0.138563\\
70	0.537755\\
71	0.235293\\
72	0.336838\\
73	2.131631\\
74	2.897183\\
75	2.777012\\
76	2.896760\\
77	2.811029\\
78	2.852447\\
79	2.682585\\
80	2.849636\\
81	2.958019\\
82	3.009394\\
83	2.889143\\
84	2.904556\\
85	3.083849\\
86	3.168144\\
87	2.979217\\
88	2.869782\\
89	3.051994\\
90	3.262383\\
91	3.316503\\
92	3.284719\\
93	3.273750\\
94	3.505389\\
95	3.262338\\
96	3.471251\\
97	3.733316\\
98	3.613280\\
99	3.808387\\
100	3.811436\\
};
\addlegendentry{Latent Tracking without Temporal Correlation}

\addplot[
    line width=0.8pt,
    color=profRed,
    mark=*,
    mark repeat=6,
    mark size=1.4pt
]table[row sep=crcr]{%
1	-1.674900\\
2	2.295817\\
3	-1.932829\\
4	-4.353256\\
5	-3.503608\\
6	0.890393\\
7	-2.667046\\
8	-0.956206\\
9	-0.148030\\
10	-2.533773\\
11	-3.048245\\
12	-0.325214\\
13	-1.921763\\
14	-2.618001\\
15	-1.269111\\
16	-2.631953\\
17	-3.468077\\
18	-3.768583\\
19	-2.382060\\
20	-1.415207\\
21	-2.251508\\
22	-2.453587\\
23	-2.097865\\
24	-3.350105\\
25	-1.788724\\
26	-2.852348\\
27	-2.005240\\
28	-1.046548\\
29	-1.165125\\
30	-1.098125\\
31	-1.590016\\
32	-1.120095\\
33	-0.911498\\
34	-1.112243\\
35	-0.950671\\
36	-3.087347\\
37	-4.246845\\
38	-5.396770\\
39	-4.884897\\
40	-5.272074\\
41	-3.655181\\
42	-0.070388\\
43	0.091565\\
44	-2.289505\\
45	-3.652350\\
46	-2.746713\\
47	-3.646649\\
48	-5.077120\\
49	-3.987301\\
50	-3.418615\\
51	-2.926222\\
52	-3.115308\\
53	-6.380140\\
54	-2.967375\\
55	-4.913346\\
56	-5.567457\\
57	-7.883551\\
58	-6.129896\\
59	-4.495529\\
60	-5.236466\\
61	-5.354229\\
62	-5.960511\\
63	-3.305909\\
64	-3.701979\\
65	-5.880681\\
66	-7.350989\\
67	-6.615383\\
68	-4.904879\\
69	-6.995910\\
70	-6.713386\\
71	-3.795203\\
72	-5.435851\\
73	-4.654892\\
74	-4.959881\\
75	-4.562402\\
76	-6.701540\\
77	-7.428493\\
78	-9.575022\\
79	-9.399738\\
80	-8.974024\\
81	-10.189881\\
82	-5.672887\\
83	-7.454691\\
84	-7.695272\\
85	-8.092739\\
86	-6.417754\\
87	-7.814346\\
88	-7.797095\\
89	-6.059252\\
90	-7.377096\\
91	-4.512467\\
92	-7.290771\\
93	-6.825940\\
94	-4.276972\\
95	-6.266649\\
96	-3.360437\\
97	-6.700144\\
98	-7.250422\\
99	-5.903731\\
100	-5.556371\\
};
\addlegendentry{Latent Tracking with Temporal Correlation}

\addplot[
    line width=1.1pt,
    densely dotted,
    color=profPurple
]table[row sep=crcr]{%
1	-0.114628\\
2	1.473912\\
3	-0.710854\\
4	-3.012615\\
5	-2.308821\\
6	0.540987\\
7	-1.903284\\
8	-0.583456\\
9	0.983717\\
10	0.228695\\
11	-0.264410\\
12	0.713556\\
13	-2.432880\\
14	0.354470\\
15	-1.918762\\
16	-2.100665\\
17	-2.922263\\
18	-1.846506\\
19	-0.102097\\
20	0.407275\\
21	0.134666\\
22	0.357798\\
23	-1.090166\\
24	0.492669\\
25	0.479192\\
26	-0.183899\\
27	0.208857\\
28	-1.255937\\
29	0.809142\\
30	-0.733538\\
31	-0.179535\\
32	-1.987652\\
33	0.099219\\
34	0.778255\\
35	0.022712\\
36	-2.311374\\
37	-4.695666\\
38	-4.888973\\
39	-4.419097\\
40	-3.332135\\
41	-2.827143\\
42	-0.578906\\
43	-0.670710\\
44	-2.847735\\
45	-2.535894\\
46	-1.906364\\
47	-1.739952\\
48	-2.877162\\
49	-2.300235\\
50	-0.422680\\
51	-2.541228\\
52	-3.244054\\
53	-2.603574\\
54	-1.233582\\
55	-1.038793\\
56	-1.170569\\
57	-0.852845\\
58	-0.541534\\
59	0.271932\\
60	0.412500\\
61	0.182986\\
62	0.499581\\
63	-0.048175\\
64	0.301351\\
65	0.206700\\
66	-0.409436\\
67	-0.731106\\
68	-1.191832\\
69	-4.647800\\
70	-3.918375\\
71	-2.702848\\
72	-3.669220\\
73	-0.118071\\
74	1.167066\\
75	-2.957137\\
76	-4.747941\\
77	-3.885125\\
78	-5.450808\\
79	-8.938571\\
80	-11.121973\\
81	-9.982281\\
82	-10.132299\\
83	-10.212283\\
84	-8.769229\\
85	-11.870196\\
86	-5.652071\\
87	-8.042354\\
88	-10.033696\\
89	-4.784910\\
90	-4.175076\\
91	-7.771653\\
92	-5.319883\\
93	-6.109467\\
94	-6.544441\\
95	-6.535993\\
96	-6.037486\\
97	-5.770957\\
98	-7.338992\\
99	-6.421437\\
100	-6.989059\\
};
\addlegendentry{Direct Channel Tracking Benchmark}

\end{axis}
\end{tikzpicture}
    \vspace{-2.5em}
    \caption{Channel Estimation NMSE Over Time for $20\times 20$ BS.}
    \label{fig_a400}
\end{figure}
Figure~\ref{fig_a100} presents the overall channel estimation performance of the proposed method over time. The results demonstrate that the proposed method consistently outperforms the traditional \ac{LS} estimator. Moreover, as expected, without the temporal correlation constraint, the \ac{LSTM} network fails to effectively track the latent dynamics, resulting in significantly higher estimation errors. We also compare the proposed method with an end-to-end tracking approach that directly estimates the full channel matrix from received pilot signals using \ac{LSTM} networks, without employing a pretrained latent representation. While this end-to-end method achieves comparable or superior performance due to its higher degrees of freedom in training, it becomes increasingly challenging to train as the number of antennas grows, since this deep \ac{LSTM} network complexity must scale accordingly, whereas the \ac{LSTM} network in the proposed method does not. To demonstrate this scalability advantage, we increase the number of \ac{BS} antennas to $20\times 20 = 400$ while maintaining all other system parameters. The results in Fig.~\ref{fig_a400} reveal that the direct end-to-end \ac{LSTM} tracking method experiences a performance degradation, as the expanded network size requires significantly more training data to achieve sufficient convergence. In contrast, the proposed method maintains more robust performance across different antenna configurations, clearly demonstrating its superior scalability and effectiveness for massive \ac{MIMO} systems.

\section{Conclusion}
This paper presents a novel deep learning approach for massive \ac{MIMO} channel estimation that combines autoencoders and \ac{LSTM} networks to exploit temporal correlation. The method learns low-dimensional channel representations and tracks them across coherence intervals using limited pilots. The key contributions include a training methodology that preserves temporal correlation in the latent space and a decomposed architecture that separates channel encoding from dynamic tracking, enhancing its scalability for large-scale systems.

\bibliography{references}
\bibliographystyle{IEEEtran}

\end{document}